\documentclass[useAMS,usenatbib]{mn2e}

\usepackage{psfig}

\def\H0{{$H_0$}}

\def\imm#1{\raise0.4pt\hbox{$\langle$}$#1$\raise0.4pt\hbox{$\rangle$}}
\def\Imm#1{{\raise0.4pt\hbox{$\langle$}{#1}\raise0.4pt\hbox{$\rangle$}}}
\def\Square#1{{\raise0.4pt\hbox{$[$}{#1}\raise0.4pt\hbox{$]$}}}
\def\log#1{${\rm log}(#1)$}

\def\lsim{{\small\mathrel{\hbox{\rlap{\hbox{\lower2pt\hbox{$\sim$}}}\raise2pt\hbox{$<$}}}}}
\def\gsim{{\small\mathrel{\hbox{\rlap{\hbox{\lower2pt\hbox{$\sim$}}}\raise2pt\hbox{$>$}}}}}

\newcommand{\apj}{ApJ}

\newcommand{\mnras}{MNRAS}
\newcommand{\aap}{A\&A}

\begin{document}

\title
[Cepheid parameters by template-fitting]
{Determination of Cepheid parameters by light-curve template-fitting}
\author[N. R. Tanvir et al.]
       {N. R. Tanvir$^{1}$, M. A. Hendry$^{2}$,
        A. Watkins$^{1}$, S. M. Kanbur$^{3}$,
\newauthor L. N. Berdnikov$^{4}$ and C. C. Ngeow$^{5}$\\
       $^{1}$Centre for Astrophysics Research, University of Hertfordshire,
       College Lane, Hatfield, Herts. AL10 9AB. UK.\\
       $^{2}$Department of Physics and Astronomy, University of Glasgow, G12 8QQ,
       Glasgow, UK.\\
       $^{3}$Department of Physics, State University of New York,
       Oswego, NY 13126. USA.\\
       $^{4}$Sternberg Astronomical Institute, Universitetskij Prospekt 13, 
       Moscow 119992. Russia.\\
       $^{5}$Department of Astronomy, University of Illinois, 1002 W Green St.,
       Urbana-Champaign, IL 61801. USA.}

\date{Accepted .
      Received ;
      in original form }

\pagerange{\pageref{firstpage}--\pageref{lastpage}}

\maketitle

\label{firstpage}

\begin{abstract}
We
describe techniques to characterise the light-curves of regular
variable stars by applying principal component analysis (PCA) to a training
set of high quality data, and to fit the resulting light-curve
templates to sparse and noisy photometry to obtain parameters such
as periods, mean magnitudes etc. The PCA approach allows us
to efficiently represent the multi-band light-curve shapes of each
variable, and hence quantitatively describe the average behaviour of
the sample as a smoothly varying function of period, and also the
range of variation around this average.

In this paper we focus particularly on the utility of such methods for
analysing HST Cepheid photometry, and present simulations which
illustrate the advantages of our PCA template-fitting approach.  These
are: accurate parameter determination, including light-curve shape
information; simultaneous fitting to multiple passbands; quantitative
error analysis; objective rejection of variables with non Cepheid-like
light-curves or those with potential period aliases.

We also use PCA to confirm that Cepheid light-curve shapes are
systematically different (at the same period) between the Milky Way
(MW) and the Large and Small Magellanic Clouds (LMC, SMC), and consider
whether light-curve shape might therefore be used to estimate the mean
metallicities of Cepheid samples, thus allowing metallicity corrections to be
applied to derived distance estimates.
\end{abstract}


\section{Introduction}

Many astrophysical investigations rely on the determination of parameters
of periodic variable stars.  Notably, the use of Cepheid variables as 
distance indicators requires estimation of periods and (usually) 
intensity-mean magnitudes in order to establish a period--apparent
luminosity relation.
With sparse and noisy data this is hard to do reliably.
Given the large investment of HST time in observations of Cepheids in
nearby galaxies (eg. Freedman et al. 2001; Saha et al. 2001;
Tanvir et al. 1995), it is particularly important for the techniques
employed to be as accurate and efficient as possible.

A number of algorithms have been developed to 
objectively estimate variable star
parameters. Notably the ``string length'' method of Lafler and Kinman 
(1965), which essentially minimises square magnitude differences
between successive phased data-points, is still frequently used to determine 
periods. This method works well, especially with precise and well-sampled data,
but is likely to be less secure with data ``at the limit'' -- i.e. close
to the limiting apparent magnitude of the photometry and/or with sparse
phase coverage. To find intensity-mean magnitudes, many authors use the 
phase-weighted method suggested by Saha \& Hoessel (1990), which makes 
allowance for the non-uniform sampling of the light-curve in time.
Again, this works well with good data, but is potentially
inefficient (in the sense of not making full use of all the data)
with sparsely-sampled data.

Most HST studies have gone one step further in using the shape of the
light-curve in the $V$ band to predict its form in the $I$ band, and
hence to allow the $I$ band intensity-mean magnitude to be estimated
from only a very few photometric data-points.
The motivation for this approach is to provide colour information 
relatively cheaply, which is required to estimate -- and
then correct for -- reddening by dust.

The simplest such recipe (Freedman 1988) uses only prior 
knowledge of the typical ratio of $V$ to $I$ band amplitude and the
typical phase shift between $V$ and $I$ bands at maximum-light for 
Cepheids. With this model, a correction can be made to the 
$I$ mean photometry, assuming that it is the same as the correction 
which would have to be applied to an equivalently undersampled $V$ light 
curve, multiplied by the adopted ratio of amplitudes. Obviously, as
with other similar methods, errors 
are introduced here, both those dependent on the $V$ and $I$
photometric quality (or lack thereof) 
and possibly also the accuracy of the prior information.
Subsequently a rather more sophisticated algorithm was developed
by Labhardt, Sandage \& Tammann 1997. This
involves predicting and fitting a template light-curve in the $I$ 
band based on the parameters (i.e. the period, phase, amplitude and shape) 
already determined from the $V$ band data.  The strong correlations 
between the light-curves of Cepheids in different bands make this a 
productive approach.

Fitting template light curves as a means of estimating Cepheid
parameters was first introduced by
Stetson (1996) who used
templates based on Fourier 
decomposition of a set of well-observed MW and LMC/SMC Cepheids.
In his method, initial values of plausible periods
are determined by string-length analysis, and then templates 
fitted with each of these periods as a starting point,
and the overall amplitude left as a free parameter (in
addition to the period, phase, and mean magnitudes).  
A scoring system is then used to identify the most plausible
fit.
Stetson argued that the advantage of automated classification
of variables and determination of their parameters is not so
much that a computer algorithm will necessarily do better than
an experienced human analyst, but that the biases and
systematics can be more easily studied and characterised.

A further refinement to the Fourier-fitting method was presented in
Ngeow et al. (2003), where ``simulated annealing'' is
used to improve the quality of the Fourier decomposition of 
sparsely-sampled HST $V$ band light-curves. 
This technique restricts the allowed range for
the Fourier amplitudes in the minimisation procedure, and thus performs
substantially better than conventional least-squares fitting on data with
significant gaps in phase coverage.
$I$ band light curves are reconstructed from the $V$ band using interrelations
of the Fourier coefficients.

A new approach to Cepheid light-curve template generation was introduced by 
Tanvir et al. (1999; see also Hendry et al. 1999) who used Principal 
Component Analysis (PCA) to statistically characterise a training set of 
MW, LMC and SMC Cepheids, and fitted these templates to $V$ and $I$ data 
for HST observed Cepheids in M96.
By fitting well-defined and realistic template curves, several parameters
can be determined, together with estimates of their uncertainties.
One of the very attractive features of this technique is that
photometry in different bands can be handled simultaneously, so that the 
natural correlations between bands are automatically built into the 
templates and all of the data is used to determine the parameters.
Kanbur et al. (2002) described the PCA method in more detail,
used it to investigate variation in Cepheid light-curve structure
as a function of period, and described the error properties of
the PCA coefficients.  PCA template-fitting was also successfully 
applied to HST-observed Cepheids in NGC1637 by Leonard et al. (2003).

In this paper we provide a complete description of the PCA-based method 
of characterising light-curves, present an updated training set 
and consider in detail the subsequent template-fitting algorithm
which was used in Tanvir et al. (1999) and Leonard et al. (2003).
We describe
simulations which illustrate the potential of the methods,
and discuss future directions.
Although we focus on their application to $V$ and $I$ band 
Cepheid data, these techniques
may easily be extended to other passbands and
also used to analyse other classes of periodic variable stars.
For example, Kanbur and Mariani (2004) consider PCA of 
photometric data for RRab stars.

The structure of the paper is as follows. In Section 2 and 3 we 
present our training set of well-observed MW and LMC Cepheids
and describe in  more detail how PCA is applied 
in order to define the template light-curves, and the advantages
of this approach over other methods.
In section 4 we discuss
an algorithm for fitting templates to noisy data in order to estimate
light-curve parameters and their errors.  Of course, such a procedure is
required however the templates are generated, but in our case the fitting
process also returns estimates for the coefficients of the first two principal
components. In Section 5 we then go on to generate simulations of 
poorly-sampled 
Cepheids with noisy photometry, mimicking ``typical'' and ``difficult'' HST
data sets, and extract their parameters by template-fitting.
We consider distance determination and light-curve 
parameter estimation using both {\em mean}- and {\em maximum}-light 
estimates. 
This serves 
to illustrate how our method performs 
in practice
compared to other methods, and also allows us to explore the limits
of HST-like data-sets. 
In section 6 we introduce an SMC Cepheid sample, and 
consider the question of whether light-curve shape, 
for either individual Cepheids or averaged over 
populations, contains other useful 
information -- in particular its potential as an indicator of
metallicity. Our conclusions are given in 
section 7.

\section{Principal Component Analysis of our Training Set}

PCA is a widely used statistical tool and has been applied in recent 
years to a number of astrophysical problems, such as spectral
classification, photometric redshift determination and morphological
analysis of galaxy  surveys (eg. Li, Kong and Cheng, 2001). 
For a detailed account of the 
statistical basis of PCA the reader is referred to e.g. Morrison (1967). 
The central principle behind PCA is easily stated, however: it provides 
a means of transforming a multidimensional dataset consisting of a 
number of statistically dependent variables into a set of statistically 
independent variables, which are the principal components. Specifically,
the first principal component is determined to be the linear combination
of the original variables which accounts for as much of the variability in 
the data as possible; the second principal component is the linear combination 
which accounts for as much of the remaining variability as possible --
subject to the constraint that it is orthogonal to the first principal
component -- and so on. In many situations the first few principal components
may explain a high proportion of the variability in the data, so that one
may substantially reduce the number of variables used to describe the data
set with very little loss of information.

\begin{figure}
\centerline{\psfig{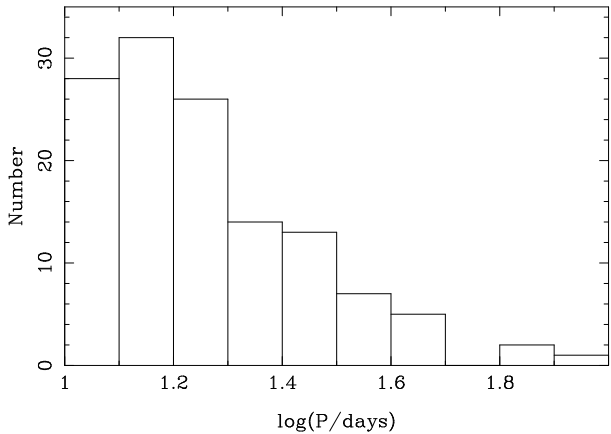}}
\caption{Distribution of log period for our `training set' of 127
Cepheids.}
\label{fig:hist}

\end{figure}

Our starting point is
 a calibrating set of 127 Cepheids, with periods $P>10$~days 
and high-quality 
well-sampled $V$ band and $I$ band light-curves. We used this `training set' 
to establish relationships between multicolour light-curve shape
(LCS) and period. 
The training set consists of:
\begin{itemize}
\item $61$  Galactic Cepheids with
photometry from Berdnikov (unpublished data-base) ,
Berdnikov \& Turner (1995),
and Moffett \& Barnes (1984)
\item $66$  LMC Cepheids, covering a wider period range
with photometry
primarily from the OGLE catalogue of
Classical Cepheids (Udalski et al. 1999a; web archive at
http://bulge.princeton.edu/$\sim$ogle/ogle2/cep\_lmc.html; Fourier
analysis from Ngeow et al. 2003), but
supplemented by data taken from various sources, particularly
 Moffett et al. (1998).
\end{itemize}
Figure \ref{fig:hist} shows the distribution of periods in our sample. Note 
that we do not include any SMC Cepheids in our training set; this reflects 
the fact that the metallicity of target galaxies for e.g. HST Cepheid 
distance estimation is generally significantly higher than that of the SMC.
We do consider SMC Cepheids in Section 6, however, in our discussion of light-curve 
shape characterised by PCA as a possible diagnostic of metallicity.

To apply PCA to this sample, we first Fourier analyse the photometric data
(for those variables not already analysed by Ngeow et al.), 
up to eighth order. ie. perform a least-squares fit of the following:

\begin{equation}
m(t)=m_0+\sum_{k=1}^{k=8}a_k~{\rm sin}(2\pi kt/P)+b_k~{\rm cos}(2\pi kt/P)
\end{equation}

The coefficients of the Fourier terms constitute
a vector consisting of 32 elements for each member of the training set
(i.e. 8 sine amplitudes and 8 cosine amplitudes for both the $V$ and
$I$ bands -- but note that the phase is shifted such that the first cosine
term in $V$ is always zero.  Note also that the mean $V$ and $I$ magnitudes,
the $a_0$ terms, 
of the calibrating Cepheids are not included in the PCA since they are 
distance dependent). 

The mean $V$ and $I$ light-curve shape is established
simply by averaging these vectors. 
PCA is then applied to the whole set of residual vectors (ie. with 
the average vector subtracted) in order to determine 
the most significant variations from the mean LCS. 
Full numerical details of the analysis are given in 
appendix A.
Incorporating both $V$ and $I$ data in each vector means that
the correlations between the coefficients in each band are
automatically encoded in the resulting analysis.
This could, of course, be extended to more bands, but we restrict
ourselves to $V$ and $I$ here since only those filters have been
used in the large majority of HST studies.

In practice, 
the first principal component largely reflects simple variations in amplitude. 
Subsequent components encode more subtle light-curve shape information,
such as ``bumps''. 
Of course, sets of 
Fourier amplitudes are not the only vectors which could be used as 
input to the PCA. One could, for example, work directly with 
the observed $V$ and $I$ magnitudes for each calibrator, smoothed and 
interpolated onto a regular grid of phase values. We find, however, that
the use of Fourier components as input vectors works very well, naturally
incorporating a degree of smoothing of the input data and providing 
a link with previous approaches to LCS analysis.

Table 1
shows for our calibrating set the proportion of the 
variance explained by the first few principal components. We can see from 
this table that one requires only a few components to explain a 
large proportion: for example, the first three principal 
components account for 89 per cent of the variance of LCS
within the sample. 
Moreover, since the observed scatter in the data 
includes the effects of photometric errors and finite sampling on
the estimated Fourier coefficients which are input to the PCA, the proportion
of the {\em intrinsic} variation in 
LCS explained by the first three
principal components will, in fact, be even higher than 89 per cent.

\begin{table}
\begin{tabular}[t]{ccc}
Component & Normalised variance & Cumulative variance \\ \hline
    1     &  0.627             & 0.627 \\
    2     &  0.199             & 0.827 \\
    3     &  0.064             & 0.890 \\
    4     &  0.026             & 0.917 \\
\end{tabular}
\label{tab:pca}
\caption{Proportion of the total variance in the calibrating set explained by
the first few principal components. (In PCA the variance associated with
each component is equal to the corresponding eigenvalue of the covariance
matrix).}
\end{table}

\begin{figure*}
\centerline{\psfig{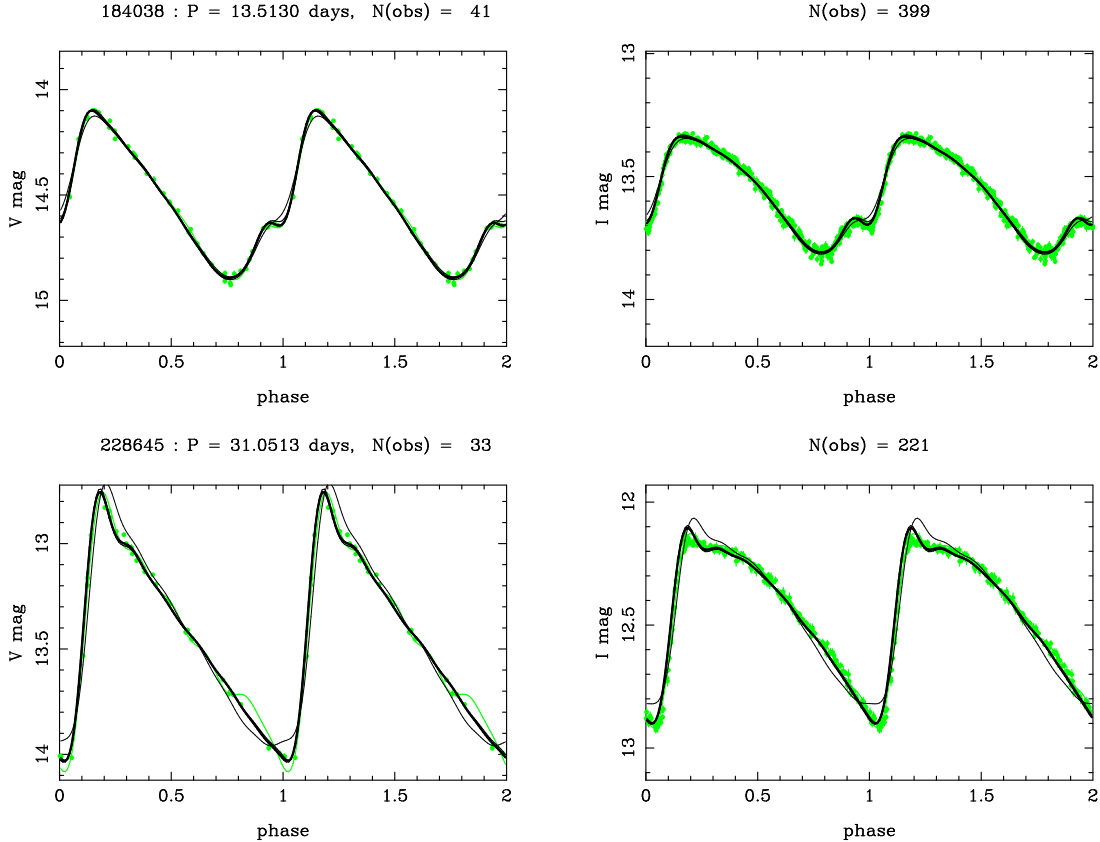}}
 \caption{Light-curves for two LMC Cepheids of different periods
(OGLE data).  The 
Fourier fits are shown as a grey
line (ie. 32 terms describing both $V$ and $I$), 
and the reconstructed PCA curves with 1, 2, 3 and 4 terms 
are shown in successively
thicker, black lines.  
We emphasise that $V$ and $I$ are simultaneously described
by the PCA curves since the analysis is performed on the combined
data-set.
Note that the fits, whilst not perfect, are
very good, and that in fact beyond 2 terms further changes in the PCA
curves are almost entirely within the thickness of the line.}
\label{fig:example}
\end{figure*}

Figure \ref{fig:example} shows an example of two Cepheids from the
OGLE data-set with very good phase coverage.  This illustrates
that excellent light-curves are reconstructed from just 2 PCA terms.
Since these reconstructed curves incorporate information from the
whole training set, they necessarily reflect average Cepheid behaviour,
and don't fit perfectly any individual Cepheid.  However, this has the
advantage that they don't follow noise in the data either, as
the Fourier fits in the $V$ band are beginning to do in these examples.

%====================================================================%

\section {Deriving Template Light-Curves}

\begin{figure*}
\centerline{\psfig{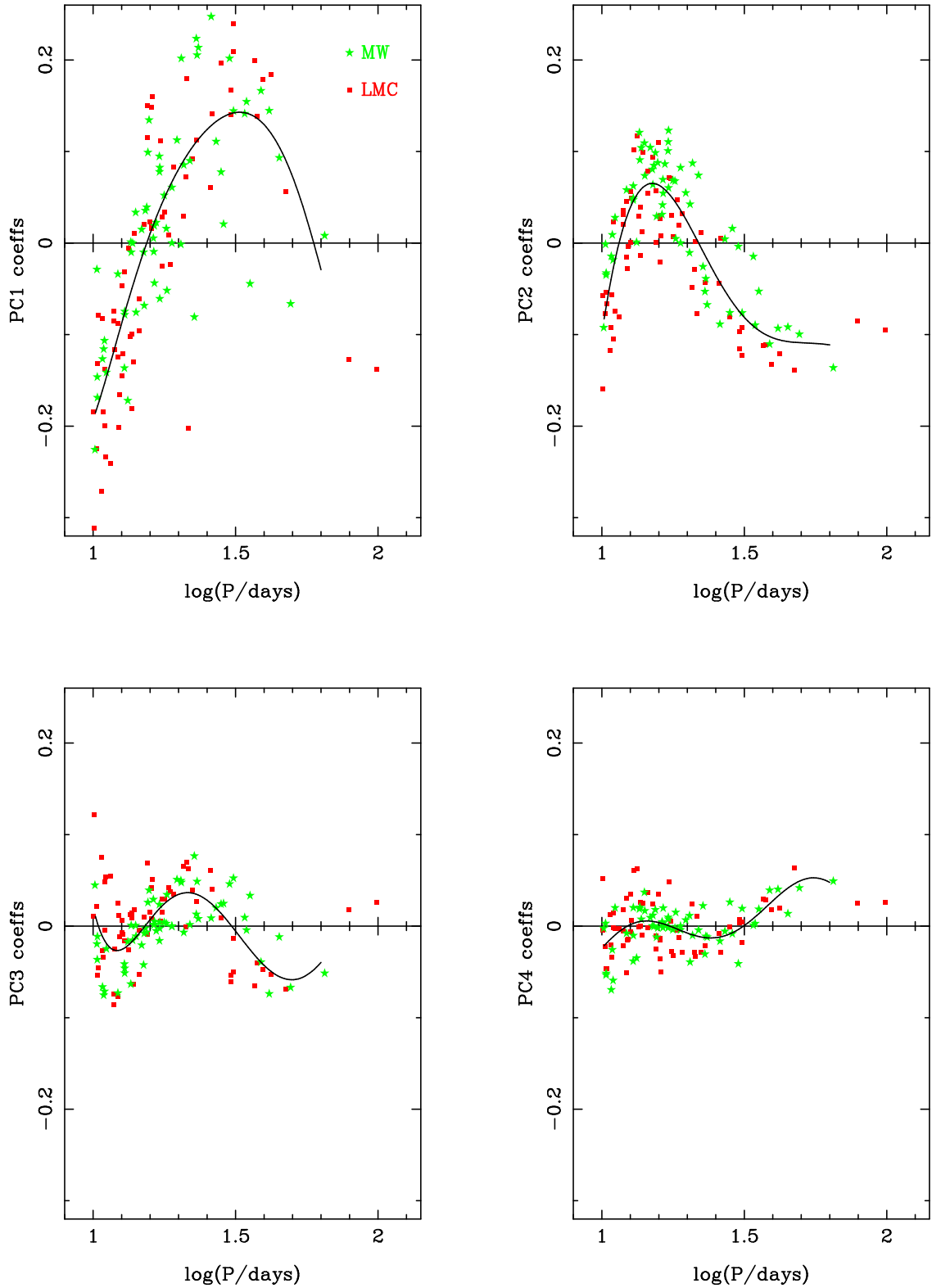}}
\caption{PCA coefficients plotted as a function of log period in days, for our 
training set of 127 Milky Way (light stars) and LMC Cepheids (darker squares).
The four panels illustrate the diminishing strength of successive principal components,
and
by plotting against log$(P)$ also reveal the systematic trends in light
curve amplitude and shape with period.  
Low order polynomial fits are overplotted which will be used to
obtain typical coefficients at a given period.
The small but real difference between the
distributions of LMC and MW points is discussed further in the text.}
\label{fig:pca}
\end{figure*}

With the PCA coefficients for each variable in hand, we can plot
them as a function of period, as shown for the first four principal
components in Figure \ref{fig:pca}. This figure reveals some
important trends -- notably that the behaviour of the first principal 
component, as expected, is similar  to 
a simple plot of amplitude versus period, with a peak
at around $P=30$ days (see e.g. Schaltenbrand \& Tammann 
1972). There is clearly also systematic
structure in the plots for the coefficients of the second and third 
principal components. 
By the fourth component the distribution of coefficients is becoming 
increasingly dominated by noise, 
although small but statistically significant correlations
of the coefficients with ${\rm log}(P)$
are seen up to at least 8 PCA terms.

We have fitted low-order polynomials through these scatter 
plots, to define ``typical'' values of the PCA coefficients 
for a Cepheid of given period, and also obtain some estimate of the spread 
around these typical values. 
The polynomial fits are shown by the solid curves in  
each of the panels of Figure \ref{fig:pca}, terminating at 
log$(P/{\rm days})=1.8$
where the number of training Cepheids becomes very few.

Before discussing the construction of template light-curves, it is
interesting to compare the distribution of coefficients for the Milky
Way and LMC subsamples.  For the first and third principal components
we can see from Figure \ref{fig:pca} that the distributions appear to
be well mixed, with no obvious distinction between the two samples.
This is broadly consistent with the results of Kanbur and Ngeow
(2004), who obtained period-colour and amplitude-colour relations for
a similar sample of MW and LMC OGLE Cepheids. While those authors
found strong evidence for a difference in the slope of these relations
between long- (i.e. $P>10$ days) and short-period Cepheids, they did
not find a statistically significant difference in the long-period
sample slopes between the LMC and MW.  For the second principal
component, on the other hand, at a given period the LMC coefficients
typically appear to be less than those for the Milky Way.  The
distributions are clearly not disjoint, so that the adoption of a
single polynomial fit to describe the structure in the combined
training set is reasonable. Nevertheless, Figure \ref{fig:pca}
suggests that the second principal component, at least, might be a
useful discriminator between different Cepheid samples.  We discuss
this point further in Section 6 below.

In order to generate a realistic Cepheid light-curve template 
all that is now required is to read off the PCA coefficients corresponding
to the desired period according to the polynomial fits in Figure \ref{fig:pca},
and hence, with knowledge of the PCA vectors and the average light-curves, 
reconstruct a full sequence of Fourier terms.
We emphasise again that because $V$ and $I$ Fourier coefficients are
both included in each vector, the light-curves in each band are
reconstructed simultaneously. 
An extra degree of sophistication can be achieved by considering the
PCA coefficients within a range around the polynomial fit corresponding
to the scatter in the training set.  In this case we find not a single
template at a given period, but a whole family of allowable
light-curves.

Recalling that the primary motivation of the present study is to 
extract optimal light-curve parameters from 
observed noisy data-sets, the next challenge is to find the
best fitting template light-curves to such a data-set when the period
and other parameters are {\it \`a priori} unknown. 
Our solution to this problem is described in more detail 
in the next section.

%=====================================================================%

\section {Light-Curve Parameter Estimation via Template-Fitting}

To characterise the Cepheid light-curves of sparsely-sampled, noisy
data our approach is to find the 
best-fit PCA templates, simultaneously in $V$ and $I$,
determined from a
search of the plausible parameter space in period, $V$ and $I$ mean magnitude, 
phase and PCA coefficients.  
Determining period, phase etc. using both bands is unusual, 
but makes most efficient use of all the data.

In practice -- mainly for computational
expediency -- only the first two 
PCA coefficients are allowed to vary.
%, and in some cases we further simplify the fitting
%procedure by fixing the second coefficient to equal its mean value for 
%the period being tested. 
The higher PCA coefficients are set 
to zero, but as seen above (Figure \ref{fig:example}), using more than two PCA terms
only modifies the light-curves at a subtle level.

The fitting procedure for an individual Cepheid is summarised as follows.

\begin{enumerate}
\item{\tt Loop over a large range of trial periods (usually between 10 and 65 days with steps
of 0.001 in the log)}
\item{\tt for each trial period, loop over a range of PC1 and PC2 coefficients around their 
typical values for that period, thus generating template light-curves in
$V$ and $I$ for each pair of values of the PC1 and PC2 coefficients.}
\item{\tt for each pair of templates, find the values of phase and intensity-mean
magnitudes in V and I which minimise the $\chi^2$ statistic, where
the $\chi^2$ is
defined as the sum of the squared deviations (normalised by the
photometric errors) between the observed $V$ and
$I$ data-points and the  magnitudes predicted by the templates.
This procedure utilizes the Amoeba algorithm (eg. Press et al. 1992).}
%\item{\tt hence obtain, for this trial period and light-curve shape (determined
%by PC1 and PC2 coefficients), the best-fit phase and mean $V$
%and $I$ band magnitudes simultaneously, by minimising $\chi^2$.}
\item{\tt move to next pair of PC1 and PC2 coefficients.}
\item{\tt move to next trial period.}
\end{enumerate}

\noindent
{\tt At the end of this process,
the trial period and $V$ and $I$ 
light-curves with the overall lowest $\chi^2$ 
is then assumed to provide the best estimates of all
the parameters.}
\vspace{5mm}

In practice we plot $\exp(-{\chi^2}_{\rm red}/2)$, as an indicator of relative 
likelihood, against test period. This plot reveals: firstly, the best period 
(the peak position); 
secondly, the goodness of the best fit (the height of the peak); 
thirdly, how well the period is determined (the width of the peak);
and, fourthly, whether there are any potential 
aliases at completely different periods (essentially the height of the
second highest peak). 
As we show later, this information can be very useful in deciding
objectively which variables to include and which
to exclude in a period--luminosity analysis.

\subsection{Estimating the distance modulus and its error}
\label{sec:dmest}

To estimate the distance modulus of the Cepheid
we now ignore the data-points and
calculate intensity-mean magnitudes, \imm{V}, \imm{I}, 
of the fitted  templates themselves.
These are compared to the absolute magnitudes calculated
for the given period using the calibrating PL relations.
The colour-excess is used to correct for extinction
following the procedure detailed by Tanvir (1997), and hence
an estimate of the true, unreddened distance modulus, $\mu_0$, is made.

We can also obtain in a straightforward manner an estimate of the
uncertainty on $\mu_0$ by computing the
posterior distribution, $p(\mu_0 | {\rm data})$, of $\mu_0$ given the
observed $V$ band and $I$ band data. Note that this procedure
accounts for internal errors due to photometric
noise and finite sampling, but not to errors in the original
templates (likely to be relatively small in practice) or
calibration errors (which must be estimated independently). 

We proceed as follows:
Let  $\phi$, $P$, ${\rm PC_1}$, ${\rm PC_2}$
denote respectively the 
phase constant, period and the first and second
PCA components of the underlying light-curves. 
Ignoring the higher order PCA coefficients, and also
neglecting for the moment the impact on the light-curve shape of other
stellar parameters such as metallicity (see below), we assume that these 
four parameters, together with the intensity-mean magnitudes, \imm{V}
and \imm{I}, completely specify the $V$ and $I$ band light-curves. 
Note, moreover, that under this approximation $P$ is uniquely defined by 
the values of \imm{V}, \imm{I} and $\mu_0$, so that we need not 
consider the period as an independent parameter\footnote{In other words 
we assume that the Cepheid lies exactly on the 
fiducial $V$ and $I$ linear PL relations}. To simplify notation,
we denote the remaining light-curve parameters collectively by the
column vector ${\bf{\Lambda}}$; i.e.
\begin{equation}
{\bf \Lambda} \equiv \left ( \Imm{V}, \Imm{I}, \phi,
{\rm PC1}, {\rm PC2} \right ) ^T
\end{equation} 

Formally, we may then write
\begin{equation}
p(\mu_0 | {\rm data} ) = \int
p( \mu_0, {\bf \Lambda} |
{\rm data} ) d {\bf \Lambda}
\label{eq:err1}
\end{equation}
i.e. we marginalise
$p( \mu_0, {\bf \Lambda} |
{\rm data} )$ over the other independent parameters. 

To simplify matters we assume that
${\rm PC_1}$ and 
${\rm PC_2}$ are equal to the values 
determined previously for the globally best-fitting template.
In practice, we also assume that 
$p( \mu_0, {\bf \Lambda} | {\rm data} ) = 0$ unless
$\phi$ is equal to its best fit estimate
for the period determined by the particular values of
$\mu_0$, \imm{V} and \imm{I}
This allows eq. \ref{eq:err1} to
be rewritten as

\begin{equation}
p(\mu_0 | {\rm data} ) = \int
p( \mu_0, {\bf \Lambda} |
{\rm data} ) d \Imm{V} d \Imm{I}
\label{eq:err2}
\end{equation}
which, in turn, can be approximated by a sum over a series of `trial'
values of \imm{V} and \imm{I}. From Bayes' theorem we may write
\begin{equation}
p( \mu_0, {\bf \Lambda} | {\rm data} ) = 
p( {\rm data} | \mu_0, {\bf \Lambda} )
p( \mu_0, {\bf \Lambda})
\label{eq:err3}
\end{equation}
where the first term is the likelihood function, expressing the probability
of obtaining the observed photometric data, given a set of light-curve
parameters and a Cepheid at distance modulus $\mu_0$, and the second term 
is a prior distribution for those parameters and for the distance modulus. 
Assuming a flat prior for $p( \mu_0, {\bf \Lambda} )$, equation
\ref{eq:err2} may be further reduced to
\begin{equation}
p(\mu_0 | {\rm data} ) \propto \sum_{j} \sum_{k} 
p( {\rm data} | \mu_0, \Imm{V}_j, \Imm{I}_k, \phi, 
{\rm PC_1}, {\rm PC_2} )
\label{eq:err4}
\end{equation}
where $\Imm{V}_j$ and $\Imm{I}_k$ denote a series of (equally
spaced) trial values of the mean $V$ and $I$ band magnitudes. One can, if
appropriate, easily generalise equation \ref{eq:err4} to the case of a 
non-uniform prior and a non-uniform grid of $\Imm{V}_j$ and 
$\Imm{I}_k$ values.

Assuming that the photometric errors are normally distributed, finally we 
obtain
\begin{equation}
p(\mu_0 | {\rm data} ) \propto \sum_{j} \sum_{k} \exp ( - \chi^2_{jk} )
\label{eq:err5}
\end{equation}
where $\chi^2_{jk}$ is the chi-squared obtained from comparing the observed
and predicted magnitudes, given values of $\mu$, $\Imm{V}_j$, 
$\Imm{I}_k$, $\phi$, ${\rm PC_1}$ and ${\rm PC_2}$. 
In fact, we could take
as our estimate of $\mu_0$ the value which maximises the posterior likelihood
$p(\mu_0 | {\rm data})$ -- or equivalently the value which minimises 
$\chi^2_{jk}$ -- however, in practice these values differ from those
for the individual best-fit template by only 1 or 2 hundredths of a magnitude
in distance modulus. Instead we use the fact that
$p(\mu_0 | {\rm data} )$ should be properly normalised, to
compute $\sigma_{\mu}$, the uncertainty in the estimated distance modulus
from the width of the resulting likelihood function.
Notice that this analysis gives reasonable estimates of the uncertainty
providing the estimated period itself
is not greatly in error.  If, in fact, the best
fit period is an alias then the true error on the distance modulus 
is likely to be significantly larger.

The above analysis can be readily extended to
include a metallicity dependence in the shape of the light-curve templates
and in the PL relations. In this case the period, $P$, would be determined by
the values of $\mu$, $\Imm{V}$, $\Imm{I}$ and metallicity, 
$Z$, which itself might be estimated along with the other independent parameters
by the template-fitting approach. Such an extension in practice seemed 
inappropriate for the training set considered in this paper, since it 
contained Cepheids from different metallicity environments (but see Section 6, 
below). The study of metallicity effects using PCA is straightforward in principle,
however (see Kanbur et al. 2002). Moreover, one could extend the model for
the prior distribution, $p(\mu_0, {\bf \Lambda})$, of light-curve parameters
to include a dependence on other fundamental stellar parameters, such as
effective temperature and mass, reflecting one's state of knowledge about
e.g. the width of the instability strip, initial mass function and mass
luminosity relation for Cepheids (see, for example, Kochanek 1997 for
an example of such a prior model).

%=============================================================================%
\section {Simulations of sparse and noisy light-curves}

Another use of our template light-curves is to provide the
underlying models for production of artificial Cepheid photometry.
In this section we describe simulations
designed to resemble the 
sparse and noisy photometry from typical HST Cepheid monitoring 
campaigns.
We then run the PCA template-fitting program on these data-sets
to obtain maximum likelihood estimates of
the parameters for each simulated Cepheid, and hence 
establish how accurately the input parameters are recovered.
We also compare the template-fitting results to those 
obtained from more traditional parameter estimation methods.
Although the simulations are realistic, they are not designed to 
replicate specific cases of HST-observed Cepheids in external galaxies but
rather to represent generic examples similar to those found by
most HST studies.
Moreover, we have not compared our template-fitting method with all algorithms 
which have been used to determine Cepheid light-curve parameters -- partly 
because many such studies have involved some degree of subjectivity, 
for example in selecting the Cepheids themselves, which is hard to replicate.

We chose to simulate data for Cepheids of log$(P/\rm days)=1.4$ 
(about 25 days), 
being typical of the variables observed in
the HST programs, and in the middle of the range for which 
the sampling strategy is optimised.
Phases were random, and PCA 
coefficients chosen to be as
those for real Cepheids (ie. based on the polynomial fit to 
the training set) at the period 
in question. The sampling of the light 
curves was carried out using a particular
sequence of observations based on that adopted by the HST $H_0$ Key-Project 
group 
Specifically
this meant 12 epochs of observation in $V$ and 4 epochs in $I$. 
Realistic, magnitude-dependent photometric noise was added to the data-points,
with one set of 400 simulations  with error bars on each point around 0.1
to 0.2 mag, being representative of
``typical'' HST data,
and another set of 1000 simulations representing
``difficult'', low signal-to-noise ratio ($S/N$) data with error bars more like 0.15 to
0.3 mag per data-point.  The latter set approximates the
worst-case data which have been obtained in some HST studies.

\subsection {Results of simulations}

To provide a benchmark we first analysed each simulated data-set
with the 
methods of period finding 
via string-length minimisation (Lafler and Kinman 1965) and phase-weighted
intensity-mean magnitude estimation (Saha and Hoessel 1990).
We then analysed the same synthetic data using our PCA template approach
described above. 
For the sake of brevity we henceforth refer to these as the ``OLD''
and ``NEW'' algorithms respectively, whilst clearly recognising 
these particular OLD methods are by no means the only ones
used in previous studies.  They do, however, have the benefit
of being easily and mechanically applied to the data.

\begin{figure}
\centerline{\psfig{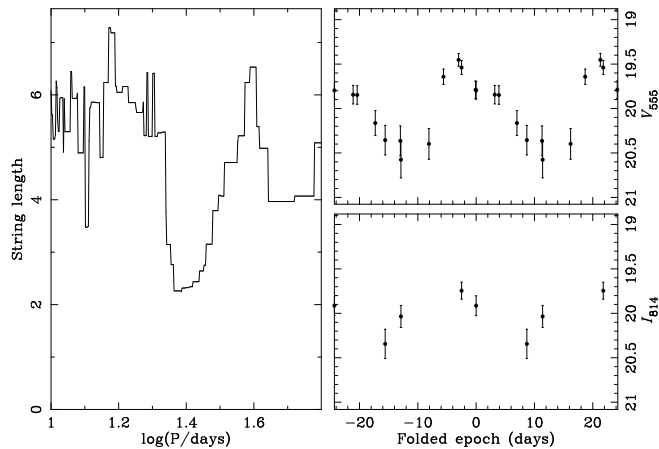}}
 \caption{Example of artificial Cepheid photometry simulated to
have $S/N$ typical of HST data and analysed
with the OLD algorithms.
The left-hand panel shows the Lafler-Kinman (1965) string-length measure 
applied just to the $V$ band data-points and plotted
as
a function of trial period -- the best period being indicated by the
minimum string length.  
Since the input period is ${\rm log}(P/{\rm days})=1.4$, the method obviously
works well for this simulation.
The worst aliases tend to occur at half the true period (ie. offset
by about 0.3 in the log), but in this case (and in fact for 
nearly all the
simulations with this $S/N$) the worst alias is not as good a fit
as is the correct period.
The right-hand panels show the data-points folded
on this best period.  Note that the zero-points for the magnitude scales
are chosen arbitrarily.}
\label{fig:oldlo}
\end{figure}

\begin{figure}
\centerline{\psfig{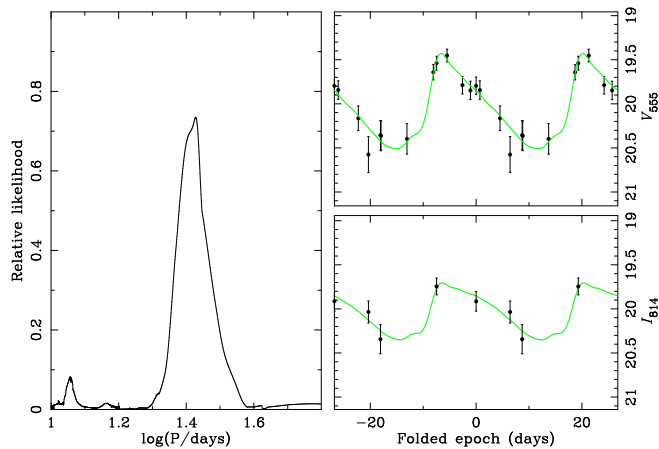}}
 \caption{The same simulated data shown in Figure \ref{fig:oldlo}
analysed with the NEW, template-fitting algorithm.
In this case we look for a peak in the relative likelihood curve 
(left-hand panel) to 
identify the best fitting period (see description in text).  
The period found by the algorithm is very close to that input
in the simulation, log($P$/{\rm days})=1.4,
and again a small alias is seen to appear near $P/2$.
The right-hand panels again show the data folded on this best period.}
\label{fig:pcalo}
\end{figure}

We consider first the simulations of ``typical'', moderate $S/N$ data.
Examples of the simulations and period determination
are given in Figure \ref{fig:oldlo} 
for the OLD
algorithms, and Figure \ref{fig:pcalo} for the PCA template-fitting algorithm.
Both approaches work well in this example,
in the sense of correctly identifying the period.
The fact that subtle light-curve shape information is largely erased by
this degree of sparse-sampling and noise addition means that
the use
of the same templates for both creating and fitting to the simulated
data should not significantly bias the results.
We checked this expectation by running further noisy simulations but
this time starting
with the observed light curves of individual Cepheids from the training
set which were outliers from the curves in Figure \ref{fig:pca}.
The results and level of improvement with template-fitting were
qualitatively similar to those reported above, with the proviso
that both algorithms (especially the OLD ones) do somewhat better 
when the overall Cepheid amplitude is larger rather than smaller.  In other
words, as one would expect, the
outliers with PC1 above the average are easier to find periods for
than the outliers with PC1 below the average.

\begin{figure}
\centerline{\psfig{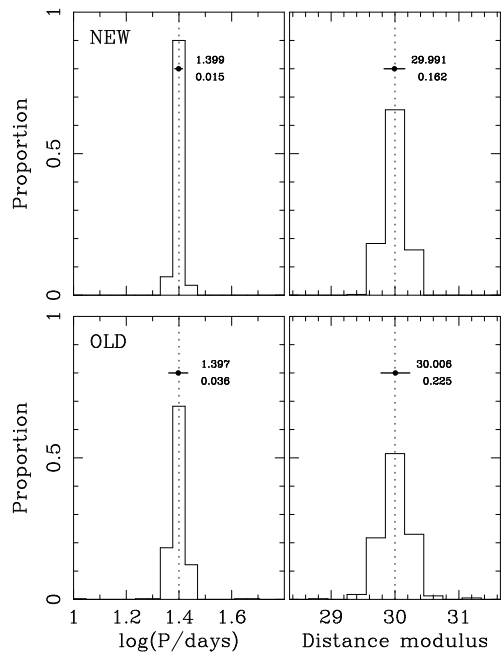}}
 \caption{Histograms of the results from 400 simulations 
with ``typical'' HST $S/N$.
The input values are indicated by vertical dotted lines, namely
${\rm log}(P/{\rm days})=1.4$ and $\mu_0=30$.  The latter is chosen
arbitrarily as being typical of HST studied galaxies.
The upper panels are for template-fitting and the lower panels using
the OLD algorithms.  Both period and distance modulus are
well determined with this $S/N$, as indicated by the solid
dot and bar which represent the mean and standard deviation of each
distribution (numerical values are printed next to the bar).  
The number of occasions where an alias period is
wrongly identified as the true period is negligible.}
\label{fig:histlo}
\end{figure}

In Figure \ref{fig:histlo} we summarise the results of all 400 simulations
in histograms showing the returned periods and inferred distance 
moduli.  The extinction-corrected
distance moduli are calculated using $P$, \imm{V} and \imm{I}
as described by Tanvir et al. (1997), which is essentially the same method
as used by the HST $H_0$ Key-Project group.
Neither method is confused by aliases with this $S/N$ data, although
the template-fitting produces rather more accurate periods and
distance moduli.  Specifically the 1$\sigma$ {\it rms} scatters 
for the NEW and OLD methods are 0.16 and
0.23 mags respectively, suggesting that the uncertainties in distance modulus
resulting from light-curve parameter estimation for
samples of several tens of Cepheids should only be a few
hundredths of a magnitude.

For each fit to the simulated data-sets we also evaluated the
uncertainty in distance modulus
as described in section 4.1.  The average turned out to be 0.15 mag,
very close to the observed scatter, giving confidence that the
errors returned by the template-fitting procedure itself
are realistic.

\begin{figure*}
\centerline{\psfig{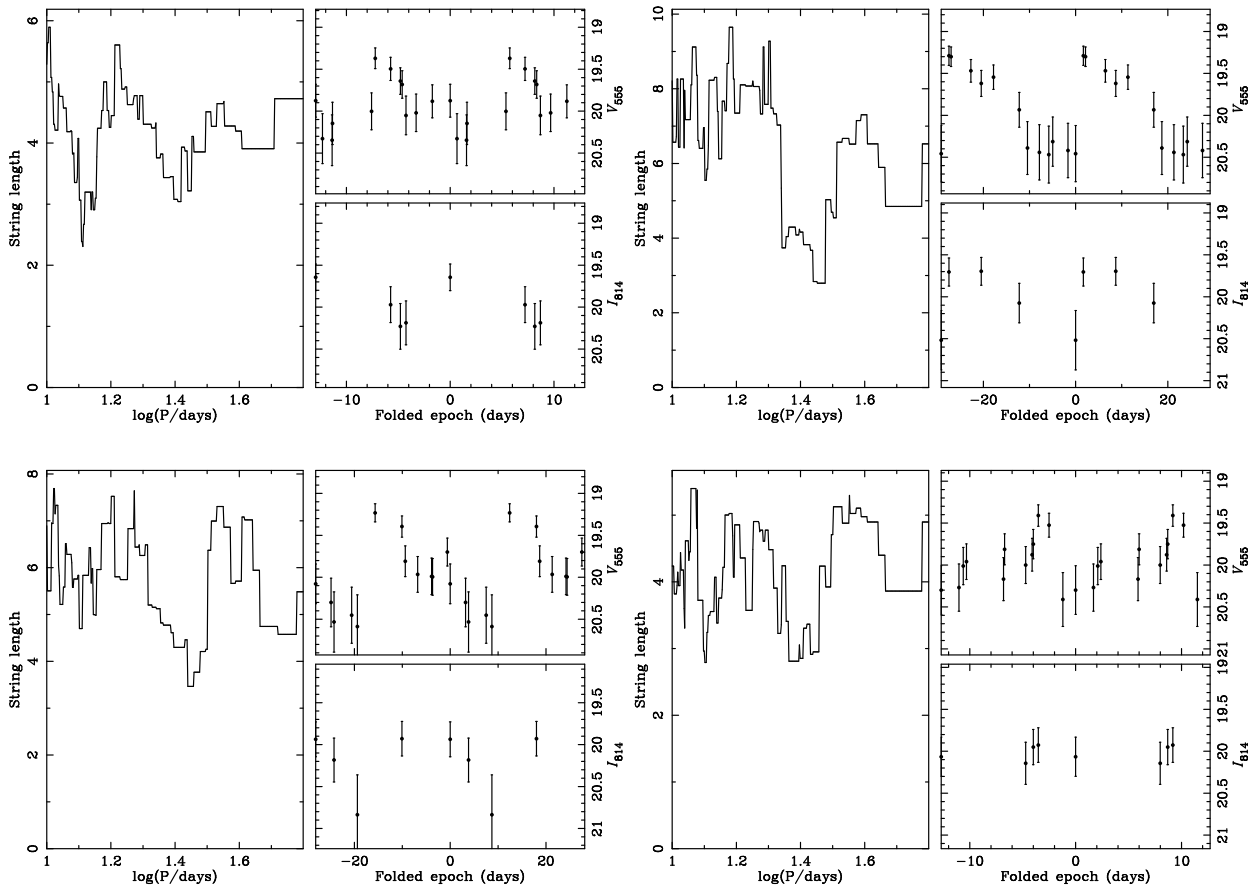}}
 \caption{Examples of ``difficult'', low $S/N$ 
simulated data analysed 
with the OLD algorithms.  
The panels are similar to those in Figure \ref{fig:oldlo}.
The four cases were chosen to illustrate
a range in behaviour, with the top left and bottom right variables
suffering from bad aliases at half the true period.
Note, in the latter case the folded data-points
do not trace a very Cepheid-like light-curve.
}
\label{fig:oldvlo}
\end{figure*}

\begin{figure*}
\centerline{\psfig{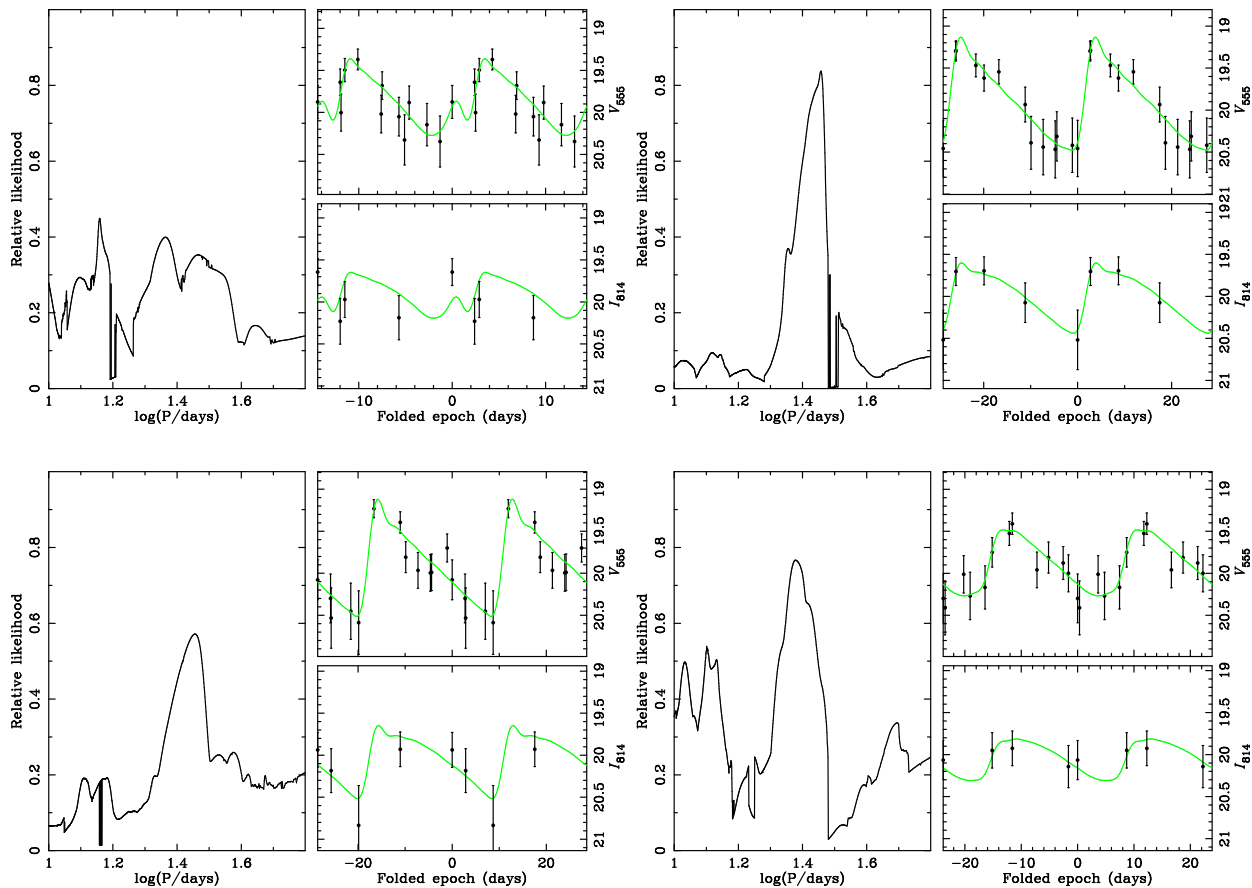}}
 \caption{The same simulated data as Figure \ref{fig:oldvlo} 
analysed with the NEW, template-fitting algorithms.  Again the top
left case hits the same problem with an alias period providing a better
fit that the true period.  However, in the bottom-right case this time,
the fact that we are fitting a template rather than simply minimising
string-length has correctly identified the period.}
\label{fig:pcavlo}
\end{figure*}

The situation with the 
``difficult'', low $S/N$ data is illustrated
in Figure \ref{fig:oldvlo} for the OLD algorithms and Figure
\ref{fig:pcavlo} for the NEW template-fitting ones.
In this case we show four examples to highlight the fact that
now results range from cases where the input parameters
are well recovered, to instances where the best 
fit is actually obtained with an alias period.

\begin{figure}
\centerline{\psfig{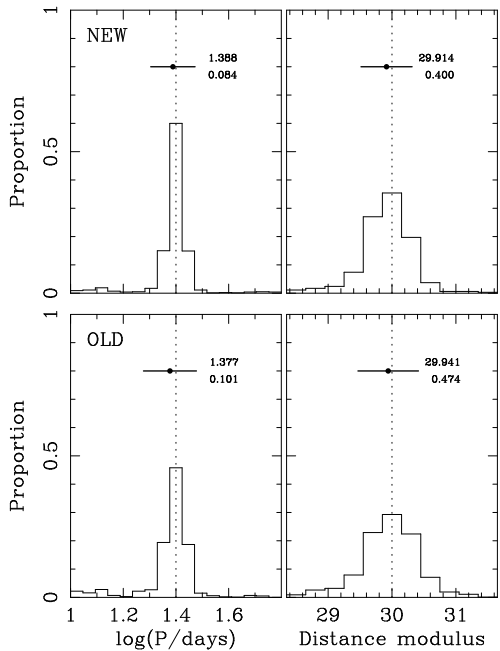}}
 \caption{Histograms of the results from 1000 simulations of
``difficult'', low $S/N$ data.  Compared to Figure \ref{fig:histlo}
we see that the distribution of returned periods and inferred distance-moduli
shows more scatter, a small but non-negligible contamination by
aliased periods and a small but statistically significant bias toward
lower values of $\mu_0$.}
\label{fig:histvlo}
\end{figure}

This behaviour is summarised in Figure \ref{fig:histvlo} 
for 1000 simulated data-sets
which shows
that now both techniques produce the occasional period 
aliases. In general
terms the periods and distance moduli show more scatter than was
the case with higher $S/N$, although overall an accuracy of
about 0.4 mag per Cepheid in distance modulus is still reasonably
good.  Once again the template-fitting performs
a little better than the OLD algorithms, but both reveal
a slight bias to lower distance moduli, over and above the increased
scatter.

However, we must be cautious in interpreting these results for
a number of reasons.  In practice, because of the low $S/N$, some of our
simulated variables would probably not have been classified
as variables in the first place had they appeared in
an HST study.  Furthermore,
bad fits would often be rejected as not being sufficiently
``Cepheid-like''.  It wouldn't be surprising if such cases
of bad template fits also produced the most discrepant distance
moduli.

In order to assess these effects, and also make the test more realistic,
we clipped
the sample of simulations to exclude those for which the degree
of scatter of the data-points about a constant, non-variable line
was such that it would only occur by chance 1 time in 5000.
In other words we insisted (as do most Cepheid studies one way
or another), that the threshold for treating a star
as a variable is high enough that very few non-variable stars
ever exceed it by chance.  Further we set an upper limit to the acceptable 
reduced ${\chi^2}_{\rm red}$ for the template fit of 1.3, which means
that 22 per cent of true Cepheids will be lost, but ensures
that only those which fold to produce genuine ``Cepheid-like'' light
curves are retained.  Finally, we rejected any variables for
which there was an alias period with a ${\chi^2}_{\rm red}<1.5$ which was
separated by less than 0.1 in log$(P)$ from the highest likelihood peak.
This procedure loses a few fits which produced accurate periods, 
but is particularly good at
removing probable aliases.

\begin{figure}
\centerline{\psfig{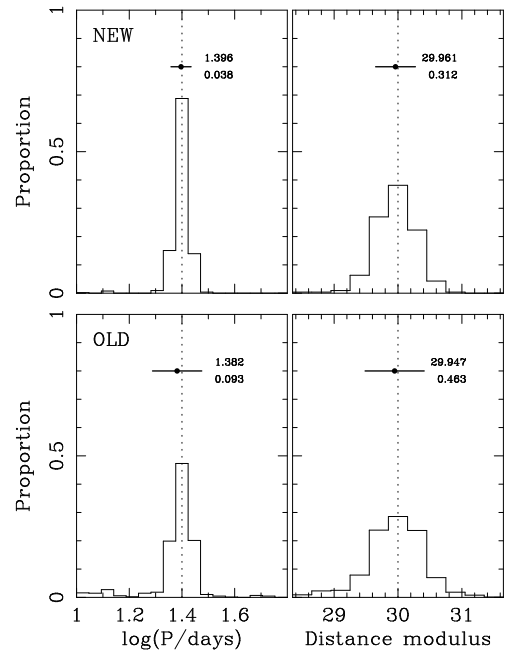}}
 \caption{The same information as in Figure \ref{fig:histvlo} 
but this time clipped of the low amplitude variables and those
with badly fitting light-curves.  This process mimics what is
often done in practice, selecting candidates above some threshold
criterion for variability, and rejecting
those which don't appear ``Cepheid-like''.
In addition, as described in the text, we have removed variables
with potential aliases, which is a fairly well-defined,
automated procedure
(ie. based on the height of the second highest peak in the relative
likelihood plot for the variable) when doing template-fitting.
These steps obviously reduce
scatter, remove many of the aliases, particularly for the template-fitting,
and largely remove the small bias.  The price which is paid is the
loss of 12.5 per cent of the variables used for the OLD algorithm analysis,
and about 46 per cent of the variables used for the template-fitting.}
\label{fig:histclip}
\end{figure}

The results of this whole clipping procedure are shown in 
Figure \ref{fig:histclip}.  As expected,  the
scatter in the results of both methods is reduced, 
but in particular problems with aliases for the 
template-fitting are largely removed, as
is the bias in distance modulus determination.

Viewed as a whole the simulations permit the following conclusions:

\begin{itemize}
  \item Template-fitting results in a roughly 30 per cent reduction
  in scatter in 
  estimates of distance modulus for the ``typical'' $S/N$ data
  compared to the OLD methods of string-length minimisation to 
  determine periods and phase-weighted averaging to obtain
  intensity-mean magnitudes.
  \item There is a tendency to slightly underestimate periods
  for ``difficult'', low $S/N$ data.
  Again template-fitting
  does somewhat better than string-length.  This small bias in period
  also leads to a small bias in distance modulus.
  In fact, we have also performed simulations for
  longer period log$(P)=1.7$ Cepheids
  (ie. around 50 days), although not reported
  here in detail, and find this underestimation becomes a little
  worse.  This is not surprising since the period
  is now approaching the total length of the observing sequence,
  and is only a little below the maximum period employed in the trials.
  \item Both approaches perform respectably even for low $S/N$ data,
  but with a increasing incidence of period aliases.
  However, not only does the template-fitting do somewhat better, 
  it also computes a goodness-of-fit ($\chi^2$) for the template fit,
  providing an objective way of selecting the variables for inclusion in the 
  analysis.  A sample culled on the basis of only including good 
  template fits reduces the scatter by a factor of around 2 compared
  to the traditional OLD methods.
  \item The distance modulus uncertainties calculated as described in
  section 4.1 are on average
  very good.  For example, for the 25 day  period simulations
  the average uncertainty calculated for the typical $S/N$ case
  was 0.15 mag compared to the actual dispersion around
  the true (input) value of 0.16 mag.  For the low $S/N$ data the
  numbers are 0.28 mag for the estimated errors compared to
  0.31 mag as the dispersion for the clipped sample of simulations.
\end{itemize}

While these are interesting results, and establish the utility of
the PCA template-fitting method, we caution that they do not
imply significant problems with the results obtained by 
the various groups reporting HST Cepheid observations in the past.  
For one thing, for the typical
$S/N$ photometry, the NEW algorithm only performed a little better
than the OLD ones.
Furthermore, most recent
studies have not simply adopted the OLD methods
considered above, but have also either applied ``chi-by-eye''
rejection of doubtful variables, and/or performed other variants on
the template-fitting scheme. The results of our simulations 
reinforce the conclusion that such template-fitting has
many benefits, but we have also shown that our procedure
has a more rigorous statistical basis and
provides a more efficient means of encoding relevant light-curve
shape information than previous methods.

\subsection {Estimating Maximum-Light from Template Fits}

Various authors have suggested that Cepheids at maximum-light may be as good,
if not better, standard candles than Cepheids at intensity-mean-light
(e.g. Sandage and Tammann 1968; Kanbur and Hendry 1996; Kanbur et al. 2003).
Aside from arguments based on intrinsic physical properties, 
another advantage could be that maximum-light may be more
precisely determined than mean-light if the Cepheid is faint
and hence poorly observed through minimum.

However, maximum-light has rarely been used in practice, perhaps partly
because many epochs are required to give a decent chance
of sampling close to the maximum.  One also loses some of the
benefit of averaging many observations to reduce noise.
Obtaining estimates of maximum-light from template fits may be
a way of benefiting from the advantages while not 
suffering the disadvantages.  All the data is used, with appropriate
weighting, and reasonable estimates of maximum-light 
can be obtained without requiring dense sampling over the maximum
itself.

We have also tested our ability to estimate maximum-light using the
template fits to the simulated data.
The resulting histograms  
are actually so similar to the ones
presented for mean-light that we feel they are not worth 
showing separately (mean-light is very marginally better).  
Admittedly
here the fact that the same templates are used to create the artificial 
noisy data
in the first place and then to fit to it, may make
the simulation results appear slightly rosier than reality.
But the point is clear: although we find no evidence that
maximum-light is superior to mean-light, it is certainly
reasonable to use maximum-light with template-fitting.
Interestingly we also find a strong correlation between the
maximum-light and mean-light distance moduli for individual
simulated Cepheids, indicating,
perhaps unsurprisingly, that they encode very similar information
and can't therefore be combined in any way to provide an improved
distance indicator.

\section {Light-Curve Shape and Metallicity}

\begin{figure*}
\centerline{\psfig{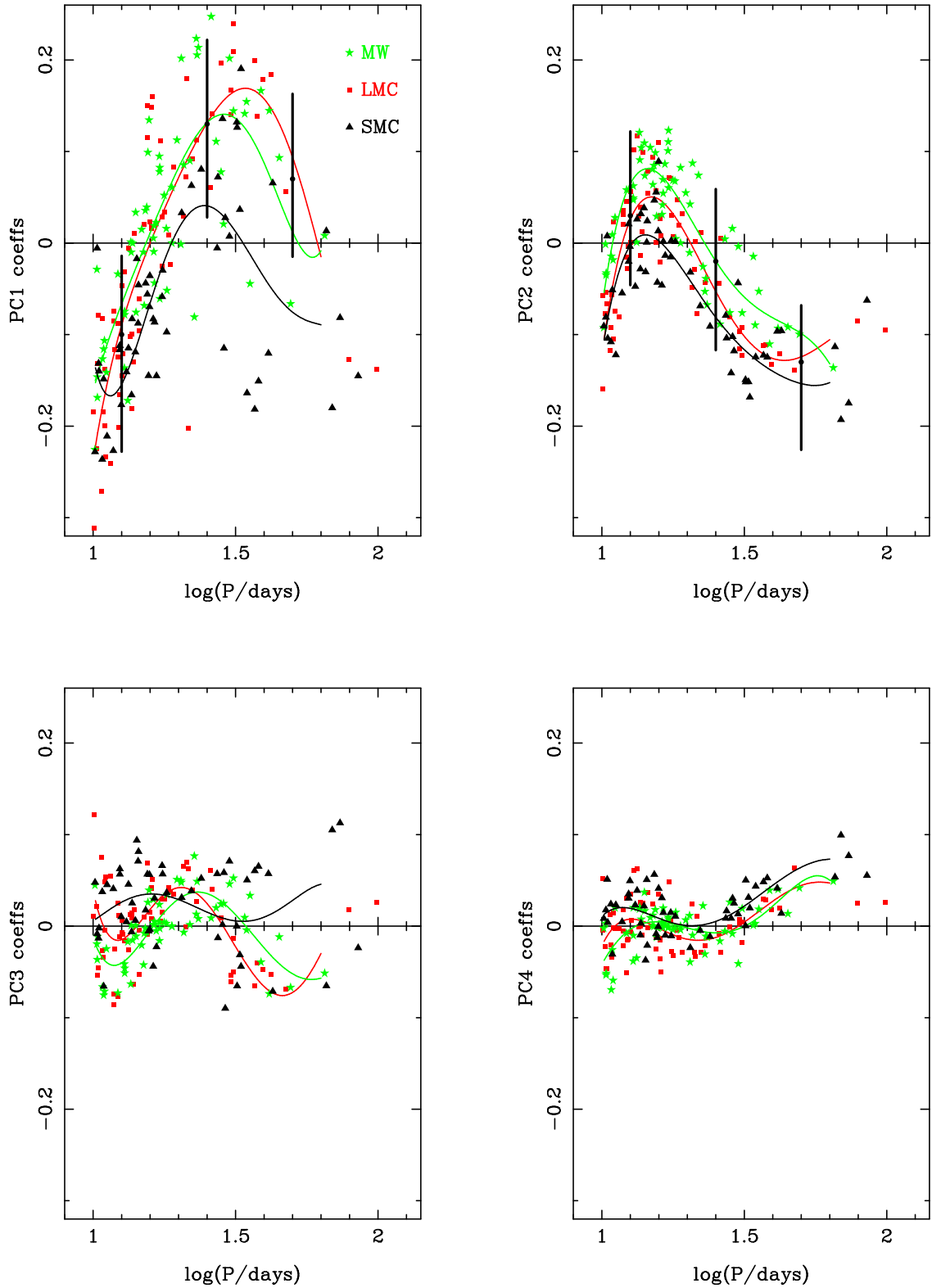}}
\caption{PCA coefficients plotted as a function of log period, for our 
training set of Milky Way (light stars), 
LMC (darker squares), extended by the addition of a sample of 
SMC (dark triangles) Cepheids.  The PCA
basis used is that of Figure \ref{fig:pca} (ie. the SMC data have been
decomposed onto this basis).  Separate low-order polynomial fits are
shown for each set of points, illustrating the systematic differences
in light-curve shape from galaxy to galaxy, presumably 
due to metallicity differences.  The large error bars
show the spread in derived  PCA coefficients expected from template-fitting
to ``typical'' 
noisy data, as estimated from our simulations.  This indicates that
photometry for individual HST-observed  Cepheids will not usually be
good enough to quantify metallicity, but averaging together a reasonable
sample for a particular galaxy might be.}
\label{fig:metal}
\end{figure*}

An obvious question which arises from our analysis thus far is whether 
light-curve shape is sensitive to physical parameters other than just 
period. It would be particularly useful, for example, if light-curve shape
were found to be sensitive to a property -- such as metallicity -- which
is expected to correlate with the absolute magnitude of a Cepheid 
(see e.g. Caputo et al. 2000, and references therein). 
Quantifying the (reddening corrected) sensitivity of Cepheid
distances to metallicity has proven hard, but estimates have 
tended to be in the region of 0.2 mags in distance modulus ($\sim$10 per cent
in distance) for a factor 10 in metallicity (eg. Sakai et al. 2004).
In 
most HST studies, the Cepheid metallicity is estimated from 
spectroscopy of the gas phase -- sometimes from nebulae in other parts of the 
galaxy from the Cepheids. Directly constraining the metallicity of the Cepheid
sample itself via observations of light-curve shape would have obvious
advantages over this approach.

In fact, Paczynski and Pindor (2000) already pointed out that 
OGLE observed Cepheids in the SMC have systematically lower
amplitudes than those in the LMC in the period range $1.1<{\rm log}(P)<1.4$,
which they suggest is likely to be a metallicity effect.
Similarly,
Kanbur et al. (2002) presented evidence suggesting a
difference in the average light-curve shape of SMC and LMC first-overtone 
Cepheids, based on a principal component analysis of densely sampled $V$ and 
$I$ Cepheid light-curves from the OGLE (Udalski et al. 1999a,b) and EROS 
(Beaulieu et al. 1995) microlensing surveys. Figure 12 of Kanbur et al. 
shows, as a function of period, the coefficients of the
first and second principal components 
for OGLE first-overtone Cepheids with periods less than 10 days. A comparison 
from that figure of the PCA coefficient
distribution for LMC and SMC Cepheids shows some 
clear differences -- most notably that the distribution of first principal 
component coefficients 
for the SMC Cepheids generally lies below that for the LMC 
Cepheids.  Hence the PCA approach picks out changes in the structure of
overtone light-curves which correlate with metallicity.
 
Figure \ref{fig:metal} shows the distribution, 
as a function of log period, of the first four 
principal component coefficients of the same MW and LMC Cepheids 
as Figure \ref{fig:pca} together with a further 
52 Cepheids from the SMC.
The data for the latter are largely from the OGLE
database (Udalski et al. 1999b) and Moffett et al. (1998).
Note that the SMC light-curves have been decomposed onto
the PCA basis established from the MW/LMC data rather than
being included in an expanded training set.

Addition of the SMC data clearly increases the spread at
a given period, but most of this appears to be systematic rather than increased
random scatter.  In particular, the PC1 coefficients are generally
lower than those of the MW and LMC, confirming the reduced
amplitude noted by Paczynski and Pindor (2000).
Particularly prominent are
a group of five SMC Cepheids with $\log(P/{\rm days})\sim1.5$,
although these
Cepheids are {\em not\/} significant outliers in the PC2 coefficient
distribution.

For the PC2 coefficients, the SMC points are again generally
lower, as is most clearly seen by looking at the separate polynomial
fits for each galaxy.
Of course, given our particular focus on HST Cepheid
studies, our training set consists only of fundamental mode Cepheids
with period, $P>10$ days. Thus Figure \ref{fig:metal} is not directly
comparable to Figure 12 of Kanbur et al. (2002).  
Unlike the plots we presented 
for the MW and LMC alone (Figure \ref{fig:pca})
we now also see that PC3 and PC4 coefficients for the SMC
tend to follow
the upper envelopes of distributions.

The metallicities of LMC and SMC Cepheids are thought to 
be about 50 per cent and 20 per cent of the MW Cepheids
respectively.
If we assume that the observed differences in
PCA coefficients are due to metallicity then it 
may provide a route to estimating the average
metallicities of other samples
of Cepheids directly.  
The feasibility of this is indicated by the bold vertical
bars in the top panels of Figure \ref{fig:metal}.
These show, at three different values of log$P$, the 1$\sigma$ spread
in returned PC1 and PC2 coefficients for the simulated
``typical $S/N$'' data.
(Given the typical $V$ and $I$ band sampling of
HST Cepheid observations, it would be unrealistic
to extract reliable light-curve shape information from the third, or
higher, principal components.) 
It is apparent that for individual Cepheids only very weak
constraints can be placed with this quality of data.  However,
with better data or by averaging a reasonable sample of 
Cepheids, a useful, direct diagnostic of metallicity may well be 
achievable.
We intend
to investigate further the dependence of PCA on metallicity in a
future paper.

\section {Conclusions}

We have presented in some detail our techniques to characterise
Cepheid light-curves using principal component analysis of
the Fourier coefficients for a set of well-observed
Cepheids in the LMC and MW.
We have also described how light-curve parameters can
be extracted by fitting these templates to sparse and noisy
data, and illustrated the method with extensive simulations.

The advantages of this approach are (i) very realistic light
curves as a (smooth) function of period are obtained with only one
or two 
principal components -- in fact they are frequently better
than the full Fourier fits to the calibrating data since
averaging over the full set of Cepheids removes some
numerical noise; (ii)
multicolour data can be accommodated with a single combined
fit which automatically accounts for the correlations between
bands; (iii) template fitting to all data (weighted by the errors
on each measurement) makes optimal use of the information
in determining light-curve parameters;
(iv) variables with poor fits (which might be produced by 
non-Cepheids or those whose photometry is badly affected by
crowding) and potential period aliases are
easily identified, and hence can be removed from consideration
by applying objective, statistical criteria (rather than, for example,
by visual inspection, as has often been the case in the past);
(v) errors can be estimated in a moderately rigorous
way, and Cepheids can be selected on the basis of goodness-of-fit; and
(vi) maximum-light is straight forward to estimate and can be used
as an alternative to mean-light in the PL relation. 

The simulations themselves show that most Cepheids observed
in HST campaigns should {\em individually} give distance moduli 
to about 0.2 mag (with the template fitting doing somewhat
better than less sophisticated approaches), indicating that
for typical sample sizes (several tens of variables), random errors
in the derived Cepheid parameters
should only be at the few per cent level
for the sample as a whole,
and systematics are likely to be the dominant source of
uncertainty.
Interestingly we have found that even with very poor $S/N$
data, errors can be as little as 0.3 mag for individual Cepheids
using template fitting.

Finally we note that these methods can easily be extended to
photometry in more than two bands, and to the analysis of other 
kinds of periodic variable stars.

\section*{Acknowledgments}

This work made use of the McMaster Cepheid Database, maintained
by Doug Welch.

\appendix

\section[]{Reconstruction of Cepheid light curves}

Although the primary purpose of this paper is to illustrate
the general advantages of template fitting in obtaining Cepheid
parameters, some readers may be interested in using the
coefficients we have determined, for example to generate
template light curves for their own use.

The light curves for all Cepheids
in both V and I are initially decomposed
into Fourier terms (equation 1).
The principal component analysis allows us to rewrite these
light curves in terms of the PC vectors ${\rm\bf P}_k$ and
an average Cepheid light curve ${\rm\bf A}$.

\begin{equation}
m(t)=m_0+{\rm\bf A}+{\sum_{k=1}^{k=32}}{\gamma_k {\rm\bf P}_k}
\end{equation}

Each PC vector (and indeed the ``average'' light curve vector) 
are simply a sum of sine/cosine terms, the coefficients for which
are given in Table A1.

\begin{equation}
{\rm\bf P}_j={\sum_{k=1}^{k=16}}{\alpha_k {\rm sin}(2\pi kt/T) +
                      \beta_k {\rm cos}(2\pi kt/T)}
\end{equation}

\begin{table*}
\begin{tabular}[t]{rrrrrrrrrrrrrrrr}

$\alpha_1$ & $\alpha_2$ & $\alpha_3$ & $\alpha_4$ & $\alpha_5$ & $\alpha_6$ & $\alpha_7$ & $\alpha_8$ & $\alpha_9$ & $\alpha_{10}$ & $\alpha_{11}$ & $\alpha_{12}$ & $\alpha_{13}$ & $\alpha_{14}$ & $\alpha_{15}$ & $\alpha_{16}$  \\
$\beta_1$ & $\beta_2$ & $\beta_3$ & $\beta_4$ & $\beta_5$ & $\beta_6$ & $\beta_7$ & $\beta_8$ & $\beta_9$ & $\beta_{10}$ & $\beta_{11}$ & $\beta_{12}$ & $\beta_{13}$ & $\beta_{14}$ & $\beta_{15}$ & $\beta_{16}$  \\ \hline
  0.395 &  0.000 & -0.018 &  0.101 & -0.022 & -0.037 &  0.030 & -0.002 & -0.004 &  0.018 & -0.009 & -0.004 &  0.004 & -0.005 &  0.000 &  0.000 \\ 
  0.245 &  0.064 & -0.015 &  0.057 & -0.011 & -0.023 &  0.019 &  0.001 & -0.004 &  0.012 & -0.007 & -0.004 &  0.003 & -0.004 &  0.001 &  0.001 \\ \hline \hline
  0.523 &  0.000 & -0.039 &  0.469 & -0.347 & -0.077 &  0.108 & -0.216 &  0.098 &  0.103 & -0.076 &  0.034 &  0.005 & -0.042 &  0.004 &  0.000 \\
  0.335 &  0.125 & -0.059 &  0.275 & -0.220 & -0.059 &  0.071 & -0.130 &  0.057 &  0.063 & -0.048 &  0.018 &  0.005 & -0.025 &  0.003 &  0.002 \\ \hline
  0.376 &  0.000 & -0.141 & -0.127 &  0.271 & -0.319 &  0.266 &  0.328 & -0.316 &  0.123 & -0.024 & -0.220 &  0.111 &  0.026 & -0.030 &  0.017 \\ 
  0.233 & -0.019 & -0.048 & -0.091 &  0.211 & -0.183 &  0.162 &  0.221 & -0.198 &  0.074 & -0.025 & -0.144 &  0.074 &  0.010 & -0.022 &  0.014 \\ \hline
 -0.383 &  0.000 & -0.134 & -0.081 & -0.255 & -0.295 &  0.389 & -0.179 &  0.015 &  0.330 & -0.233 & -0.076 &  0.062 & -0.117 &  0.010 &  0.028 \\ 
 -0.308 &  0.012 & -0.089 & -0.057 & -0.141 & -0.202 &  0.225 & -0.119 &  0.001 &  0.192 & -0.165 & -0.051 &  0.041 & -0.075 & -0.002 &  0.015 \\ \hline
 -0.256 &  0.000 &  0.138 &  0.409 & -0.224 &  0.021 &  0.082 &  0.150 & -0.333 & -0.051 &  0.278 & -0.276 &  0.055 &  0.260 & -0.109 & -0.043 \\ 
 -0.198 &  0.057 &  0.047 &  0.227 & -0.117 &  0.022 &  0.022 &  0.113 & -0.238 & -0.096 &  0.202 & -0.174 &  0.036 &  0.204 & -0.114 & -0.037 \\ \hline
 -0.071 &  0.000 & -0.658 &  0.036 & -0.075 & -0.300 & -0.148 &  0.068 &  0.004 & -0.251 &  0.148 &  0.159 & -0.167 &  0.067 & -0.047 & -0.098 \\ 
 -0.041 & -0.144 & -0.367 &  0.067 & -0.046 & -0.183 & -0.075 &  0.004 &  0.037 & -0.169 &  0.110 &  0.117 & -0.117 &  0.064 & -0.051 & -0.080 \\ \hline
 -0.065 &  0.000 & -0.397 &  0.033 &  0.131 &  0.393 & -0.178 & -0.127 & -0.074 &  0.200 & -0.173 & -0.322 &  0.308 & -0.103 & -0.103 &  0.071 \\ 
  0.097 & -0.216 & -0.239 &  0.125 & -0.020 &  0.235 & -0.154 & -0.122 & -0.058 &  0.055 & -0.047 & -0.220 &  0.175 & -0.021 & -0.089 &  0.017 \\ \hline
 -0.074 &  0.000 &  0.079 & -0.320 & -0.235 & -0.098 & -0.008 & -0.267 & -0.066 & -0.062 &  0.301 & -0.105 &  0.058 &  0.212 & -0.125 & -0.035 \\ 
  0.429 & -0.479 &  0.214 & -0.166 & -0.164 & -0.087 &  0.068 & -0.181 &  0.024 &  0.043 &  0.076 & -0.077 &  0.042 & -0.007 & -0.011 & -0.025 \\ \hline
  0.149 &  0.000 & -0.037 & -0.236 & -0.173 &  0.155 & -0.006 & -0.108 & -0.016 &  0.156 & -0.074 & -0.039 & -0.144 &  0.035 & -0.254 & -0.552 \\ 
 -0.029 &  0.326 & -0.062 & -0.232 &  0.024 &  0.070 & -0.052 &  0.013 & -0.099 &  0.011 &  0.046 & -0.069 & -0.057 &  0.118 & -0.224 & -0.419 \\ \hline
 -0.331 &  0.000 &  0.142 &  0.185 &  0.186 & -0.122 & -0.103 &  0.215 & -0.100 & -0.005 & -0.048 &  0.012 &  0.014 & -0.320 &  0.060 & -0.453 \\ 
  0.301 & -0.134 &  0.066 &  0.269 & -0.062 & -0.070 & -0.021 & -0.042 & -0.015 &  0.116 & -0.142 &  0.101 & -0.008 & -0.260 &  0.079 & -0.303 \\ \hline
 -0.037 &  0.000 &  0.147 &  0.110 &  0.190 & -0.253 & -0.096 & -0.030 &  0.357 & -0.103 & -0.129 &  0.030 &  0.397 &  0.011 & -0.441 & -0.012 \\ 
 -0.030 &  0.013 &  0.056 &  0.001 & -0.027 & -0.187 &  0.022 &  0.127 &  0.217 & -0.164 &  0.042 & -0.012 &  0.271 &  0.096 & -0.363 & -0.026 \\ \hline

\end{tabular}

\label{tab:pcavecs}
\caption{The top row in this table gives the coefficients for the
average Cepheid light curve, to be used in conjuction with A2.  The
subsequent rows are the coefficients for the first 10 PC vectors.}
\end{table*}

In order to generate typical Cepheid light curves at any
given period, the $\gamma_k$ coefficients can be obtained
from the fits to the training-set data 
shown in Figure 3.

\begin{equation}
\gamma_k=\sum_{k}\lambda_k ({\rm log}(P)-1.4)^k
\end{equation}

where period, $P$, is in days. 

The coefficients for these equations are given in Table A2,
and similarly in Table A3 for the polynomial fits shown in Figure 11.

\begin{table*}
\begin{tabular}[t]{rrrrrrrrrr}
& $\lambda_0$ & $\lambda_1$ & $\lambda_2$ & $\lambda_3$ & $\lambda_4$ & $\lambda_5$ & $\lambda_6$ && scatter \\ \hline
$\gamma_1$ & 0.124 &  0.313 & -1.201 & -0.281 & -4.546 &  -2.476 & 17.944 && 0.076 \\ 
$\gamma_2$ &-0.035 & -0.557 &  0.717 &  3.530 & -8.867 &  -0.906 & 10.359 && 0.042 \\
$\gamma_3$ & 0.028 & -0.232 & -1.439 &  3.233 &  7.009 & -14.092 &  3.154 && 0.037 \\
$\gamma_4$ &-0.013 &  0.034 &  0.986 &  0.745 & -7.507 &  -2.470 & 14.399 && 0.024 \\ \hline
\end{tabular}
\label{tab:polyfits}
\caption{
The coefficients determined in equation A3 from a polynomial fit (shown in
Figure 3)
as a function of log period to the first four principal component coefficients,
$\gamma_1,~\gamma_2,~\gamma_3,~\gamma_4$, for our training set.
The final column gives the rms scatter of 
the data points around
the fits.}
\end{table*}

\begin{table*}
\begin{tabular}[t]{rrrrrrrr}
&& $\lambda_0$ & $\lambda_1$ & $\lambda_2$ & $\lambda_3$ & $\lambda_4$ & $\lambda_5$  \\ \hline

           &LMC&  0.131 &  0.505 & -1.214 & -3.228 & -2.477 &  11.339 \\
$\gamma_1$ &MW &  0.132 &  0.301 & -2.291 & -3.442 &  5.852 &  17.230 \\
           &SMC&  0.041 & -0.054 & -2.765 &  3.913 &  11.513 & -20.621 \\
&&&&&&& \\
           &LMC& -0.053 & -0.577 &  0.588 &  3.480 & -6.082 &  1.647 \\ 
$\gamma_2$ &MW & -0.020 & -0.478 &  0.739 &  1.970 & -7.774 &  3.412 \\
           &SMC& -0.081 & -0.411 &  0.529 &  1.050 & -5.489 &  8.000 \\
&&&&&&& \\
           &LMC&  0.025 & -0.367 & -1.570 &  4.613 &  9.032 & -17.867 \\
$\gamma_3$ &MW &  0.036 & -0.102 & -1.510 &  1.718 &  6.689 & -8.885 \\
           &SMC&  0.015 & -0.136 &  0.217 &  2.044 & -0.943 & -5.493 \\
&&&&&&& \\
           &LMC& -0.012 &  0.102 &  0.774 & -0.901 & -3.864 &  5.015 \\
$\gamma_4$ &MW & -0.007 & -0.005 &  0.664 &  1.357 & -3.699 & -3.809 \\
           &SMC&  0.005 &  0.113 &  0.638 & -0.680 & -2.547 &  2.894 \\ \hline
\end{tabular}
\caption{As for table A2, but in this case describing the polynomial
fits displayed in Fig 11.}
\end{table*}

\end{document}